# Boron nitride-graphene in-plane hexagonal heterostructure in oxygen environment

E. Magnano[1,2,‡], S. Nappini[1], I. Píš[1], T.O. Menteş[3,‡], F. Genuzio[3], A. Locatelli[3] and F. Bondino[1*]

[1] IOM-CNR, Istituto Officina dei Materiali, AREA Science Park Basovizza, 34149 Trieste, Italy
[2] Department of Physics, University of Johannesburg, PO Box 524, Auckland Park 2006, South Africa
[3] Elettra− Sincrotrone Trieste, AREA Science Park, Basovizza, 34149 Trieste, Italy
[‡] These authors contributed equally

**Abstract**
Aiming to improve fabrication protocols for boron nitride and graphene (h-BNG) lateral heterostructures, we studied the growth of h-BNG thin films on platinum and their behavior in an oxygen environment. We employed a surface science approach based on advanced spectroscopy and imaging techniques to investigate the evolution of surface stoichiometry and chemical intermediates at each reaction step. During oxygen exposure at increasing temperatures, we observed progressive and subsequent intercalation of oxygen, and selective etching of graphene accompanied by the oxidation of boron.
Additionally, by exploiting the $O_2$ etching selectivity towards graphene at 250°C and repeating growth cycles, we obtained in-plane h-BNG layers with controllable compositions and vertically stacked h-BN/Gr heterostructures without the use of consecutive transfer procedures. The growth using a single precursor molecule can be beneficial for the development of versatile atomically thin layers for electronic devices.

Corresponding Author: Federica Bondino, bondino@iom.cnr.it

1. **Introduction**

Lateral in-plane heterostructure monolayers can combine the functionalities and properties of various two-dimensional (2D) materials. The lateral graphene (Gr)-hexagonal boron nitride (h-BN) heterostructure (h-BNG) is of particular interest not only because it combines the distinct electronic properties of Gr and h-BN, but also because it includes additional possible catalytically active sites for the oxygen reduction reaction through bonds at the unique interface between h-BN and Gr[1]. This 2D material has a wide range of potential applications in gas sensors and atomic layer circuits[2].

The interaction of environmental molecules with 2D materials is a central topic of current research. The performance of devices based on 2D materials can be affected by their chemical stability, particularly under ambient conditions. Oxygen, is abundant in the atmosphere and can be adsorbed and alter the properties of these materials with significant impact on their applications. Oxygen is involved in most catalytic reactions and it can modify fundamentally the chemical and physical properties of materials through oxidation, intercalation, and etching[3],[4],[5],[6],[7].

To determine the suitability of h-BNG for potential applications it is crucial to evaluate its environmental physicochemical stability. In catalytic applications, for instance, the formation of B–O bonds in h-BNG could be of interest as borates can covalently bind to Gr and act as intermediates to the O═O breaking step in the reduction process of $O_2$ to produce $H_2O$[8]. Furthermore, the stability of this material in an oxidizing environment could affect the processing and performance of the devices based on it, which could have positive or negative effects on technological applications. The linking edges between h-BN and Gr domains are predicted to be highly reactive towards dissociative adsorption of oxygen molecules, which can lead to the dissociation of the h-BN – Gr boundaries [9].

In this work we investigate the chemical and morphological changes in h-BNG/Pt(111) upon exposure to molecular oxygen gas using a surface science approach. The stability of the h-BNG layer on Pt(111) against oxygen is evaluated by high-resolution core-level photoelectron spectroscopy. The results indicate a complex relationship/competition between intercalation and etching and that the etching is selective for Gr at a certain critical temperature. To engineer the ratios and sizes of different domains for electronic device a simple approach is needed. We have exploited the etching selectivity towards Gr to prepare layers with tunable Gr/h-BN composition as well as an oxidized h-BN layer via a simple route.

2. **Materials and Methods**

Monolayers of laterally-coexisting h-BN and Gr domains (h-BNG) on clean Pt(111) were prepared by low pressure chemical vapor deposition using dimethylamine borane (DMAB) precursor as described in ref[10]. Several h-BNG samples were prepared using this procedure, all displaying identical core level line shapes and similar stoichiometry with Gr domains with predominant coverage (71±4%) and the remaining part composed of h-BN complementary domains, both up to hundreds of nanometers size/lateral dimension [11]. The growth was performed in an ultra-high vacuum (UHV) chamber connected to the synchrotron-radiation photoemission measurement chamber of the CNR-IOM beamline BACH[12] at the synchrotron light source Elettra in Trieste. Molecular oxygen gas (purity$\geq$ 99.9999% vol.) at pressures up to $10^{-6}$ Torr was introduced in the UHV measurement chamber, while, for higher pressures the oxygen exposures were performed in the load lock chamber (base p ~$4 \times 10^{-8}$ Torr). The binding energies (BE) were calibrated with respect to the Fermi edge of the Pt substrate with an accuracy better than 0.1 eV. C 1s, O 1s, Pt 4f, B 1s and N 1s core levels were measured at a photon energy of 531 eV with a total resolution of 0.15 eV, while O 1s core level was measured at 643 eV with a resolution of 0.22 eV. C 1s, O 1s, B 1s, N 1s core levels were measured at normal emission, whereas Pt 4f spectra were acquired at 60° angle with respect to the surface normal.

The XPS data were analyzed using the KOLXPD software[13]. A Shirley background was subtracted to all the XPS peaks. Voigt line shapes were used for O 1s, B 1s, and N1s peaks fitting, while Gaussian broadened Donjiac-Sunjic line shapes were used for asymmetric Pt 4f and C 1s peaks.

The spectromicroscopy characterization was carried out using the Spectroscopic PhotoEmission and Low-

Energy Electron Microscope (SPELEEM) installed at the Nanospectroscopy beamline at Elettra[14]. This instrument can be operated in low-energy electron microscopy (LEEM) or x-ray photoemission electron microscopy (XPEEM) modes, which give structural and chemical sensitivity, respectively[15]. LEEM has about 9 nm of lateral resolution, whereas in XPEEM the lateral resolution is 30 nm[16]. LEEM was used both in bright-field and dark-field modes in order to map out structural domains. The sample investigated with LEEM/XPEEM was prepared ex-situ and annealed at 250°C for 90 minutes followed by a flash annealing for ~10-30 s to 700°C in-ultra high vacuum prior to the measurements.

3. Results and Discussion

3.1 Oxygen interaction with h-BNG/Pt(111)

To detect oxygen-induced effects in a full h-BNG monolayer on Pt(111), high-energy resolution C 1s, N 1s, B 1s, O 1s and Pt 4f core level synchrotron-radiation photoemission spectra were collected from the h-BNG/Pt(111) sample before and after each exposure. As found and discussed in previous studies[10], C 1s core level spectrum in the pristine layer displays a main component at 283.9 eV from $sp^2$-hybridized carbon atoms in Gr, while B 1s and N 1s show two components each; h-BN weakly ($B_w$, $N_w$) and more strongly interacting ($B_i$, $N_i$) with the substrate, located respectively at 189.7 eV and 190.2 eV for B 1s, and 397.4 and 398 eV for N 1s originating by the slightly different chemical surrounding and different bonding with the substrate of N and B atoms as the adsorption sites change in the a moiré pattern formed by h-BN on Pt(111) [17].

Pt 4f spectrum recorded in surface sensitive conditions displays bulk and surface components at 70.9 and 70.6 eV. In previous studies no changes were observed neither in the h-BNG layer nor in the Pt(111) substrate for molecular oxygen exposures up to $10^3$ L (1 L= $10^{-6}$ Torr·s) and at temperatures between room temperature and 300°C[10][18]. In this work, we show that the surface starts changing only at much higher exposures, i.e. $4\times10^9$ L at 100°C (Fig. 1). The changes observed are a downshift of h-BN core levels [Fig. 1 (d,e)] and weak oxygen peaks appearing in the photoemission spectra around 531.0 eV and 532.4 eV [Fig. 1 (b)]. Instead, no significant changes in C 1s core-level are observed [Fig. 1 (a)]. In previous studies, it was shown that the presence of downshifts in the core level spectra of 2D overlayers can be related to intercalation connected with increased distance and decreased hybridization between the overlayer and the substrate[19],[20],[21]. Oxygen intercalation was also observed separately in layers of Gr or h-BN on Pt(111)[22] and on other metals such as Ir(111)[23] and Ru(0001)[24],[25]. The presence of a downshift in h-BN core levels suggests that intercalation under h-BN has possibly taken place, while the absence of shifts and new spectral components in C 1s indicates neither intercalation under Gr, nor formation of epoxy (O–C–O) or carbonyl groups (–C=O) which are expected to contribute around 285.6(9) eV and 287.8 eV, respectively[26]. Indeed, the 531 eV O 1s component has similar BE as intercalated oxygen in h-BN/O/Pt(111) (previously found at 530.8 eV) [27] and thus it can be associated with O atoms intercalated under h-BN. No changes are observed in Pt 4f, as oxygen atoms below h-BN are not likely in contact with platinum, but interacting with the basal plane of h-BN. The 532.4 eV O 1s component [Fig. 1 (b)] can be associated with a B 1s component at 191.2 eV [Fig. 1 (d)] and attributed to bonds between boron and oxygen formed at the domain boundaries or at defect sites inside h-BN domains[18],[28],[29].

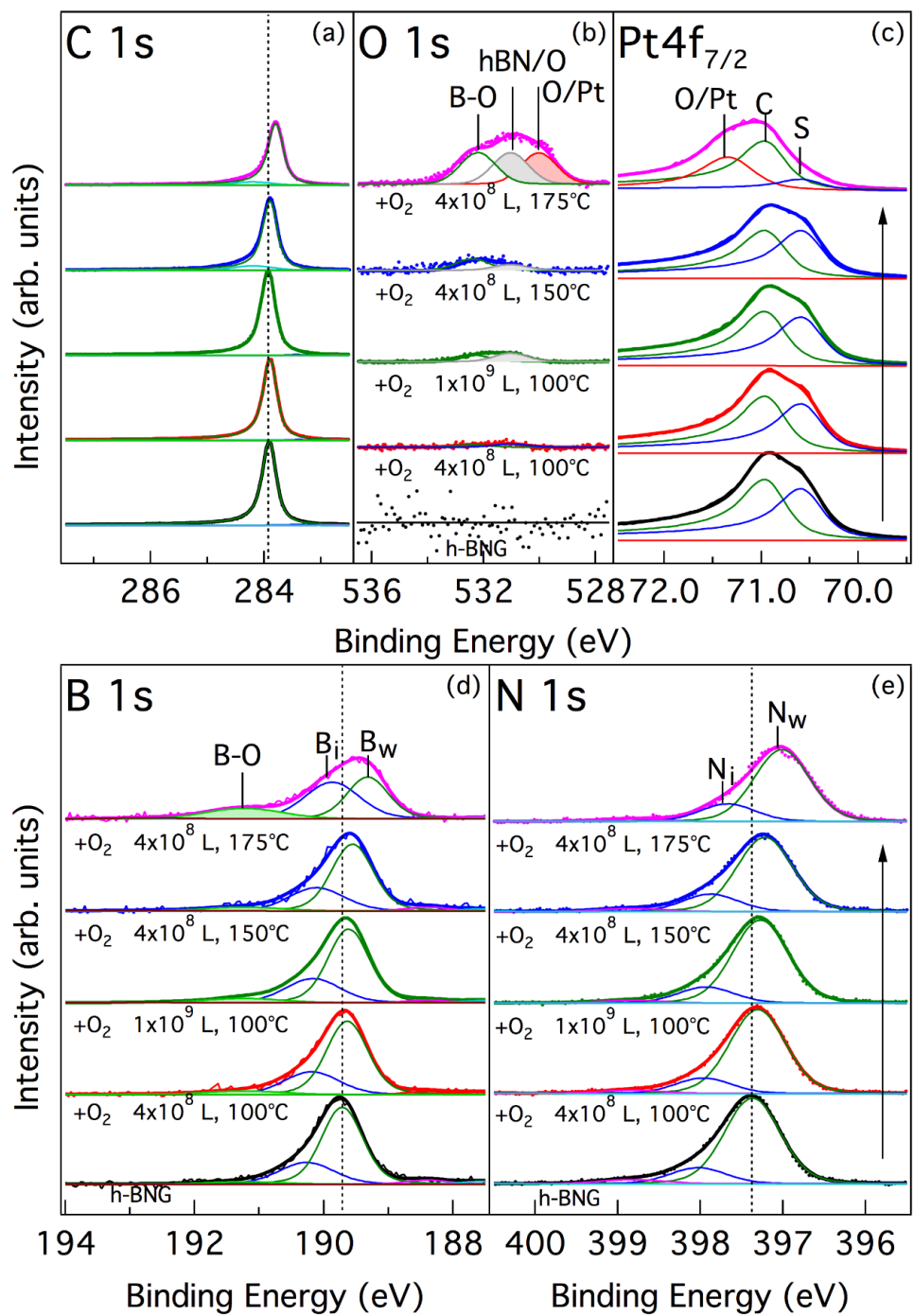

**Figure 1.** Evolution of the (a) C 1s (b) O 1s (c) Pt $4f_{7/2}$ (d) B 1s and (e) N 1s core levels of the freshly prepared h-BNG/Pt(111) layer after stepwise exposures to oxygen at different temperatures. The exposures were done in subsequent steps (from the bottom to the top, as indicated by the arrow pointing upwards), always on the same sample (the additional exposures at each step are indicated in the Figure).
The data show no changes up to 100°C when oxygen intercalates under h-BN domains, while at 175°C, oxygen intercalates also under Gr domains and more B-O bonds are formed.
The component labeled as "O/Pt" in O 1s and Pt 4f is the oxygen intercalated below Gr in h-BNG and adsorbed on platinum, C and S are the bulk and surface components in Pt 4f. $B_i$, and $B_w$ ($N_i$ and $N_w$) are strongly interacting (i) and weakly interacting (w) components in B 1s (and N 1s). h-BN/O corresponds to the component assigned to oxygen intercalated under h-BN.

The $O_2$ dissociation on h-BN is endothermic with a high energy barrier[30]. Instead, theoretical calculations predict that $O_2$ can exothermically dissociate at the B–C bridge site at the linking edges between h-BN and Gr[9].The formation of B–O bonds already at 100°C is possible due to the presence of B–C bonds at domain boundaries in h-BNG.

Further $O_2$ dose (0.32 Torr, 20 min, $4\times10^8$ L) at 175°C (Fig. 1, top curves) induces clear changes in all core level spectra. A significant downshift is found in B 1s, N 1s, but also in C 1s core levels (by 0.39 eV, 0.36 eV and 0.1 eV, respectively) with respect to the pristine layer. The Pt 4f surface component at 70.6 eV is visibly suppressed [Fig. 1 (c)] and a new component can be seen at higher BE (71.3 eV). Moreover, an additional O 1s component appears at 530 eV [Fig. 1 (b)], which is not present at lower temperatures and doses. This additional oxygen component has the BE of oxygen chemisorbed on Pt(111)[31] and the Pt 4f component at 71.3 eV is generally associated with the formation of bonds between Pt and chemisorbed oxygen[31] or the 2D overlayer[11]. Therefore, the downshift of the h-BNG core levels can be associated with oxygen intercalation between Gr in h-BNG and Pt, and the O/Pt component with oxygen chemisorbed on platinum and intercalated under Gr. In previous comparative studies of Gr or h-BN layers on Ru(0001) it was found that intercalation under Gr requires a higher temperature with respect to h-BN under similar conditions [25]. This seems to be valid also for the laterally-coexisting Gr and h-BN domains in the present case, i.e. 100-150°C are sufficient for intercalation under h-BN, while 175°C are needed for intercalation under Gr.

The O 1s and the B 1s components assigned to B–O bonds and the h-BN ($B_i$) component in B 1s have also increased. An additional small component at 284.2 eV is present in the C 1s spectrum and the total C 1s peak area displays a ~17% decrease (the decrease is starting already at 150°C) [Fig. 1 (a)]. The decrease of C appears to be linked with the increase of the B–O components in O 1s and B 1s. The 284.2 eV component can be assigned to $sp^3$/undercoordinated carbon atoms bound to Pt[11]. Importantly, according to theoretical calculations, $O_2$ can dissociate at the B–C bridge site and lead to the stretching and eventual breaking of the B–C bonds with a relatively low activation energy[9]. Thus, the changes observed at 175°C are consistent with a scenario where the opening of the h-BN-Gr linking edge takes place with the subsequent formation of B–O bonds and C dangling bonds at the linking edges which can also interact with Pt or can be easily removed. The decrease of C 1s intensity indeed suggests that some C atoms are readily removed above 150°C upon contact with oxygen as part of CO or $CO_2$.

No signature of reaction is instead observed in the N 1s core level, consistently with calculations predicting that no adsorption takes place on the C–N bonds[9]. In Fig. 2 we show a schematic model of the oxygen interaction with the hybrid h-BNG monolayer on Pt(111) at different temperatures, based on our results. Summarizing (Fig. 2 from bottom to top), no reaction between oxygen and h-BNG at room temperature; selective oxygen intercalation under h-BN domains and formation of few B-O bonds at 100-150°C; oxygen intercalation under both graphene and h-BN domains, formation of further B-O bonds and removal of few carbon atoms at 175°C.

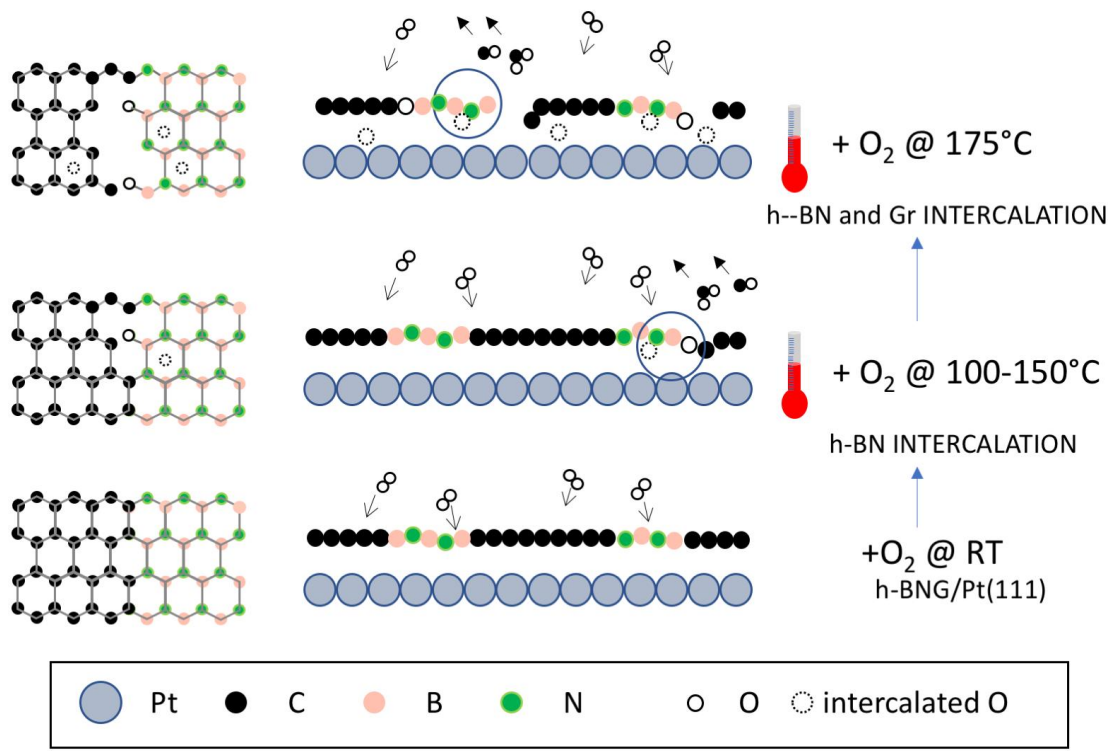

**Figure 2.** Schematic model of the oxygen interaction with h-BNG/Pt(111) (a) at different temperatures up to 175°C and (b) at a temperature of 250°C and after subsequent exposures to DMAB precursor. In each panel, left: top view, right: side view.

$O_2$ exposure (~$4\times10^8$ L) at 250°C on freshly grown h-BNG layer leads to a substantial reduction of the C 1s intensity as seen in Fig. 3. The remaining carbon presents a C 1s component at 284.2 eV, associated with Pt–C or undercoordinated carbon atoms, together with smaller C–C $sp^2$ and C–B components at 284 eV and 283.5 eV, respectively [Fig. 3 (a)]. O 1s has a main peak at 532.4 and a smaller component at 531 eV [Fig. 3 (b)]. A Pt 4f component at 0.4 eV higher BE than that of the bulk component can be seen [Fig. 3 (c)], which can be attributed to either Pt–O adsorption[31], or to the formation of Pt–C bonds[32]. Considering that O 1s peak related to oxygen chemisorbed on Pt is missing, this Pt 4f component may be associated with Pt–C bonds due to an increased hybridization between Pt and the overlayer, as it was observed after hydrogenation of graphene[32] or h-BNG [11].

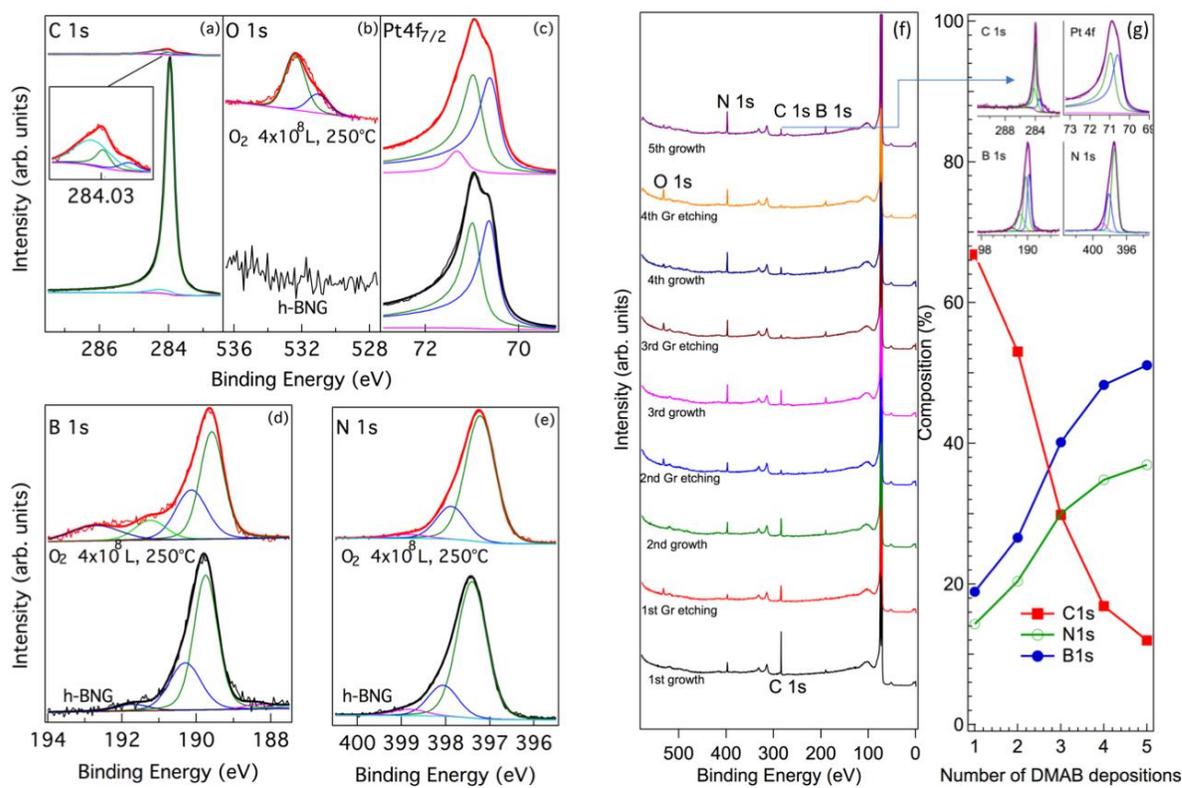

**Figure 3. (a-e)** XPS data collected from pristine h-BNG grown on Pt(111) (bottom black curves) and after Gr etching procedure which consisted in dosing $O_2$ (p=0.3 Torr, 30min at 250°C) (top red curves).
**(f-g)** XPS data collected after subsequent and repeated growth-etching cycles. XPS spectra in (f) were collected with photon energy of 643 eV from h-BNG/Pt(111) layers obtained after subsequent exposure of Pt(111) to DMAB at 730°C (first growth), full etching of graphene domains in h-BNG, exposure of the etched layer to DMAB at 730°C (second growth) etc. (g) Relative concentration of carbon, boron and nitrogen in the grown layers as obtained from the total area of C 1s, B 1s and N 1s peaks normalized to their respective photoionization cross sections. The inset in panel (g) shows the core level spectra after four etching-growth cycles.

Upon etching graphene domains with oxygen at 250°C additional B 1s components appear at 191.2 eV due to B–O bonds, and at 192.7 eV associated with O–B–B–O and O–B–O bonds[29] as seen in Fig. 3(d). N 1s [Fig. 3 (e)] does not show any sign of oxidation, but a slight (~7%) decrease in its integrated intensity. This is consistent with previous studies of h-BN oxidation, where it was found that during the formation of boron oxide O substitutes for N[27]. These results indicate that at 250°C B–C bonds are broken; some of the oxygen atoms bind to boron and platinum, and some oxygen atoms bind and reacts with carbon, desorbing likely as CO or $CO_2$ [33]. Thus, a selective etching of graphene domains in h-BNG takes place after molecular oxygen exposure at 250°C, while the majority of h-BN domains is less affected, except for a slight decrease of nitrogen, which is likely replaced by oxygen and desorbs as molecular $N_2$, and the formation of B–O bonds.

We note that the etching of graphene starts already at 150°C after dosing ~$4\times10^8$ L of $O_2$. Thus, after $O_2$ etching of Gr the surface is partially covered by h-BN and boron oxide is present (Fig. 3, red curves).

The physical-chemical atomic mechanism for gas etching of Gr has been the subject of intense investigations and these studies indicate that various parameters can have influence on it such as the temperature[21],[23], the gas partial pressure[34][27], the number of Gr layers[35], the presence of vacancies, hydroxyls, wrinkles[36],[37] point defects[33], the catalytic activity of the substrate or particles (metal catalysts such as Ru[38], Pt or Ir [33], or oxide particles such as ZnO [7]) against $O_2$ and $H_2$ [21] dissociative adsorption[38], and the role of epoxy diffusion for the initial C–C unzipping[39]. For Gr on Pt(111) the energy barrier for etching in $H_2$ is higher than in $O_2$, thus Gr etching in $H_2$ is more difficult [21]. Also h-BN can be etched away in a hydrogen or oxygen gaseous environment by Pt and other catalysts. For h-BN/Pt(111) a temperature of 450°C is required at similar $O_2$ pressures ($10^{-1}$ Torr) used in the present study [27], which is much higher compared to the temperatures considered here. Generally, both Gr and h-BN are etched anisotropically by $H_2$ and hexagonal or triangular etched holes form during etching [40][41][42].

In a previous investigation it was found that graphene can be converted to hexagonal boron nitride by a chemical conversion reaction in a quartz tube after transferring graphene on a silicon substrate [43]. An alternative strategy to increase the h-BN/Gr concentration consists of using low pressure CVD in the UHV experimental chamber with a transfer-free process, but exploiting the catalytic properties of the Pt substrate.

### 3.2 Tailoring h-BN/Gr ratio by repeated growth-etching-growth cycles of h-BNG/Pt(111)

In previous studies, nickel and platinum particle assisted etching of h-BN was used to create nano-trenches where embedded Gr nanoribbons were then grown by CVD [44],[45]. Using a related concept, we started from the Pt(111) partially covered with h-BN and h-BNO domains obtained by Gr-removal from h-BNG to grow layers with progressively increasing h-BN/Gr ratio, using a single DMAB molecular precursor.

Upon exposing the etched h-BNG to DMAB at 730°C, a new layer is obtained with higher h-BN/Gr ratio. The oxygen-related components which appear after $O_2$ etching (O 1s, B–O in B 1s and high BE component in Pt 4f) are almost fully removed after each regrowth step [Fig. 3 (f) and (g)]. The C 1s C-C $sp^2$ component around 284 eV is reduced at each etching cycle. Considering the importance of the chemical and electrical tunability of h-BNG for bandgap engineering, we have exploited consecutive growth-etching-growth cycles of h-BNG/Pt(111) using DMAB to prepare h-BNG layers with increasing amount of h-BN over Gr. Starting from a layer formed by ~67% of Gr and ~33% of h-BN after the first growth using DMAB as a precursor, followed by four etching-regrowth cycles, we obtained a layer with ~12% Gr and ~88% h-BN as estimated from XPS measurements [Fig. 3 (f)]. While the amount of boron and nitrogen are similar in the initial layer, after several etching and re-growth cycles, the amount of nitrogen slightly decreases with respect to the amount of boron, as expected [Fig. 3]. The decrease of N 1s core level intensity can be related to the formation of B–O and B–$O_2$ bonds observed in XPS after each etching (Fig. 3). In these conditions the B=N bond can be broken (likely at the boundaries with graphene), nitrogen atoms recombine and possibly desorb as $N_2$.

The recent perspective to fabricate h-BN nano-trenches with sharp and controlled edges along energy-favored crystallographic directions using metal-assisted etching [45][46] opens the possibility to fabricate nanostructures with atomic precision. This could allow trapping non-precious catalytic 3d transition metal atoms from the substrate exploiting their affinity to nitrogen atoms [47] and/or creating heterostructures with well-defined and controlled interface. This could be beneficial for tailoring 2D materials for catalysis or realizing h-BNG nano-heterostructures with etching-regrowth cycles for atomically controlled circuits [44][45].

In order to understand the evolution of the lateral heterogeneity of the h-BNG layer, the final h-BN rich layer obtained with the above procedure was characterized *ex-situ* with LEEM and XPEEM. Fig. 4 shows the LEEM and µ-LEED data for this surface. The LEEM images in Figs. 4a (at electron energy 19 eV) and 4b (7 eV) suggest that two different island-like structures coexist on the surface. In the difference image in Fig. 4c these islands, referred to as type A and type B, appear as dark and bright spots over the rest of the surface. A quantitative analysis of the island lateral distribution results in area coverages of 3% ± 1% (type A) and 20% ± 7% (type B). The largest islands reach to about 200-300 nm in diameter, whereas the average lateral extent of all islands is about 50 nm. The µ-LEED pattern from this surface, see Fig. 4d, shows the typical moiré pattern of h-BN on Pt(111), along with weak ring-like diffraction spots azimuthally rotated 30° with respect to the Pt(111) crystal directions.

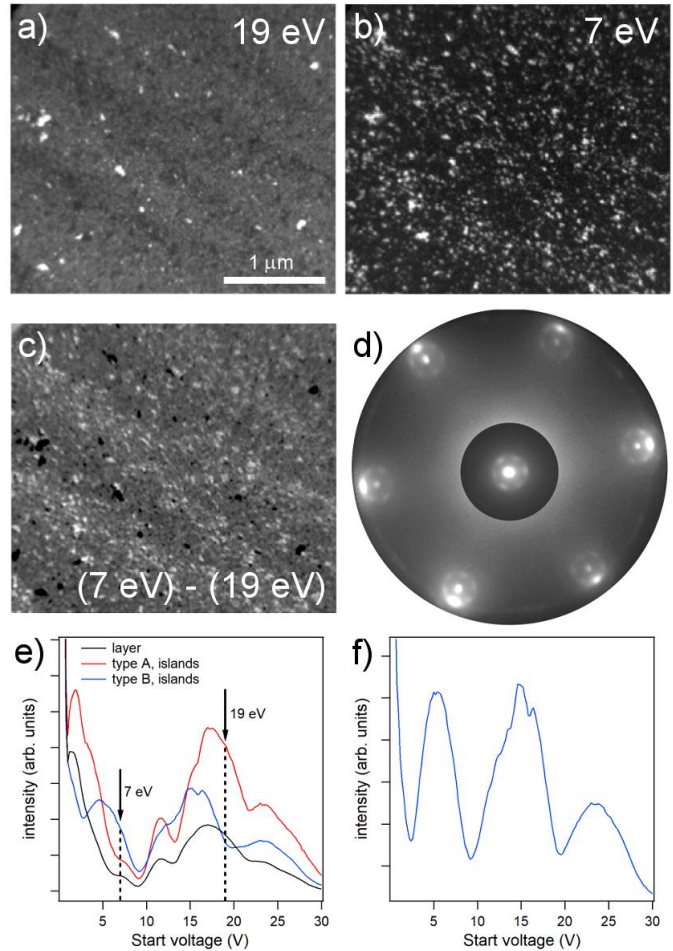

**Figure 4.** LEEM images of the layer obtained with the growth/etching/regrowth cycles at a) 19 eV, b) 7 eV. c) Difference between the two images in the first two panels. Type A (B) islands appear dark (bright) in the differential image in panel d) LEED pattern at 45 eV electron energy. The central part is shown with a different gray scale for viewing purposes. e) LEEM IV from 3 regions seen in the LEEM images. f) LEEM IV of type B islands after partially subtracting out the IV curve from the surrounding regions.

The nature of the regions giving rise to the different contrast levels in the LEEM images can be better understood by studying the respective energy-dependent low-energy electron reflectivity, i.e. LEEM I(V). The I(V) curves (Fig. 4e) are characterized by minima and maxima at specific energies, which provide a fingerprint of the film structure. Moreover, the regular quantum oscillations in the I(V) curves give information on the vertical stacking [15]. The majority of the surface (black curve) exhibits the same I(V) features of h-BN single layer on Pt(111), as can be seen by comparison to the previously published results[48]. The curves from the type A and type B islands feature additional peaks that may indicate the presence of additional layers. The curve from type A islands appears similar to the one for a monolayer h-BN/Pt(111), though with considerably higher intensity, more pronounced peaks and a few additional features (at about 4 eV and 18 eV electron energy). Nevertheless, the I(V) curve is not sufficient to fingerprint the corresponding layer configuration. On the other hand, type B islands show a distinctly different I(V) curve. Considering that the size of the islands is close to the limit of the microscope resolution, it is expected that the signal from the smallest islands is superimposed on the signal from their surroundings. Thus, by a weighted subtraction of the signal from their surroundings, the LEEM I(V) curve from type B islands reveals regular oscillations as seen in Fig. 4f. This curve shows remarkable similarity to that of h-BN/Gr bilayer on Ni(111)[49],[50].

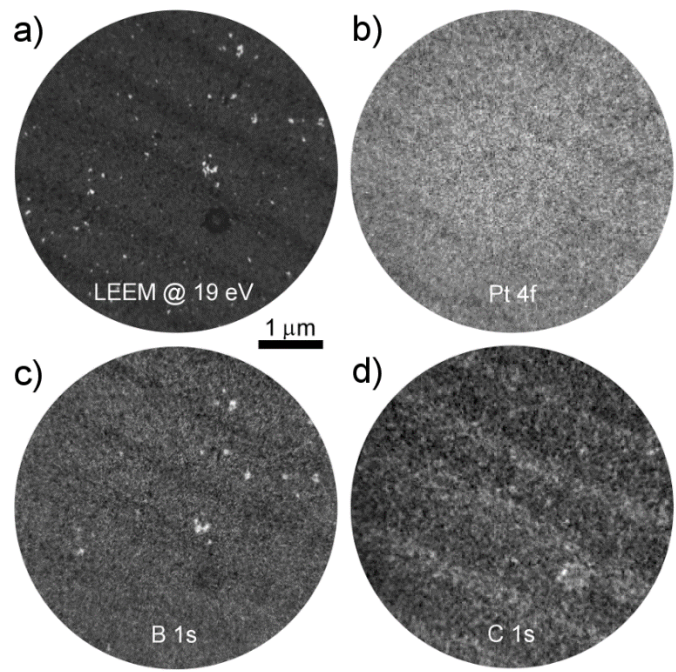

**Figure 5.** Chemical mapping of the layer obtained after etching and regrowth cycles is shown. a) LEEM image at 19 eV showing type A (bright) and type B (dark) regions. XPEEM images of the same regions at the b) Pt 4f, c) B 1s, d) C 1s core levels. Pt 4f and B 1s images are acquired with 260 eV photon energy, whereas for the C 1s image 400 eV photons are used.

Laterally-resolved XPEEM chemical maps of the heterogeneous surface are shown in Fig. 5, together with a LEEM image where type A (B) islands appear bright (dark) at 19 eV electron energy. The Pt 4f XPEEM image in Fig. 5b displays only small variations in intensity. We interpret the dark spots in the Pt 4f XPEEM image as due to the substrate signal attenuation by thicker film regions. The XPEEM images at the B 1s and C 1s core levels prove to be even more informative. The type A islands are more intense in the B 1s image in Fig. 5c. On the other hand, the type B islands appear bright in the C 1s image in Fig. 5d.

Combining the information provided by the LEEM I(V) and the XPEEM experiments, we assign the type A islands to h-BN/h-BN bilayer, whereas the type B islands to the bilayer h-BN/Gr/Pt(111). The area coverage of type B islands, 20% ± 7%, is consistent with the Gr coverage 12% found from the averaging XPS measurements as reported above. Assuming that all Gr is present in h-BN/Gr bilayer and thus covered by h-BN, the Gr coverage found from the XPS measurement should be corrected for the attenuation of the C 1s signal by the h-BN layer above. The XPS measurements were carried out at 530 eV photon energy (i.e. 240 eV kinetic energy for C 1s core level electrons). The inelastic mean free path through a single layer of h-BN can be approximated to the corresponding value for a single layer of graphite[51], and results in the attenuation of 240 eV electrons by a factor of 0.61. Correcting the 13% Gr coverage found from our XPS analysis by this attenuation factor, we estimate a Gr/h-BN ratio of about 0.24, which is in good agreement with the area coverage of type B islands as evaluated from LEEM images.

Thus, the presence of bilayer islands indicates that a confined growth[49],[52] took place in this sample. The repeated cycles of oxygen etching and regrowth lead to the formation of a h-BN layer covering a large number of graphene islands (and a minority of h-BN).

We notice that stripe-like variations are visible both in the LEEM (Fig. 4 a and b, Fig. 5 a) and XPEEM images (Fig. 5 b, c and d). These stripe-like patterns are likely due to the step morphology of the substrate. The undulations appear also in the variation of chemical species (B 1s and C 1s in Figure 5), which suggests that the step morphology could have some influence on the reactivity and intercalation.

Based on the results of our experimental data, the etching-regrowth cycles are sketched in Fig. 6. Summarizing (Fig. 6 from bottom to top), after the exposure to oxygen at 250°C (etching step) the Gr domains in h-BNG/Pt(111) layer are selectively etched away and the substrate is partially covered with h-BN islands, boron oxides and a few carbon atoms; after the exposure of this layer to di-methylamino borane at 730°C (regrowth step) a new h-BNG layer with higher h-BN/Gr with respect to the original one is obtained; finally, after several etching/regrowth steps, a h-BN layer covering Gr (and few h-BN) islands is obtained.

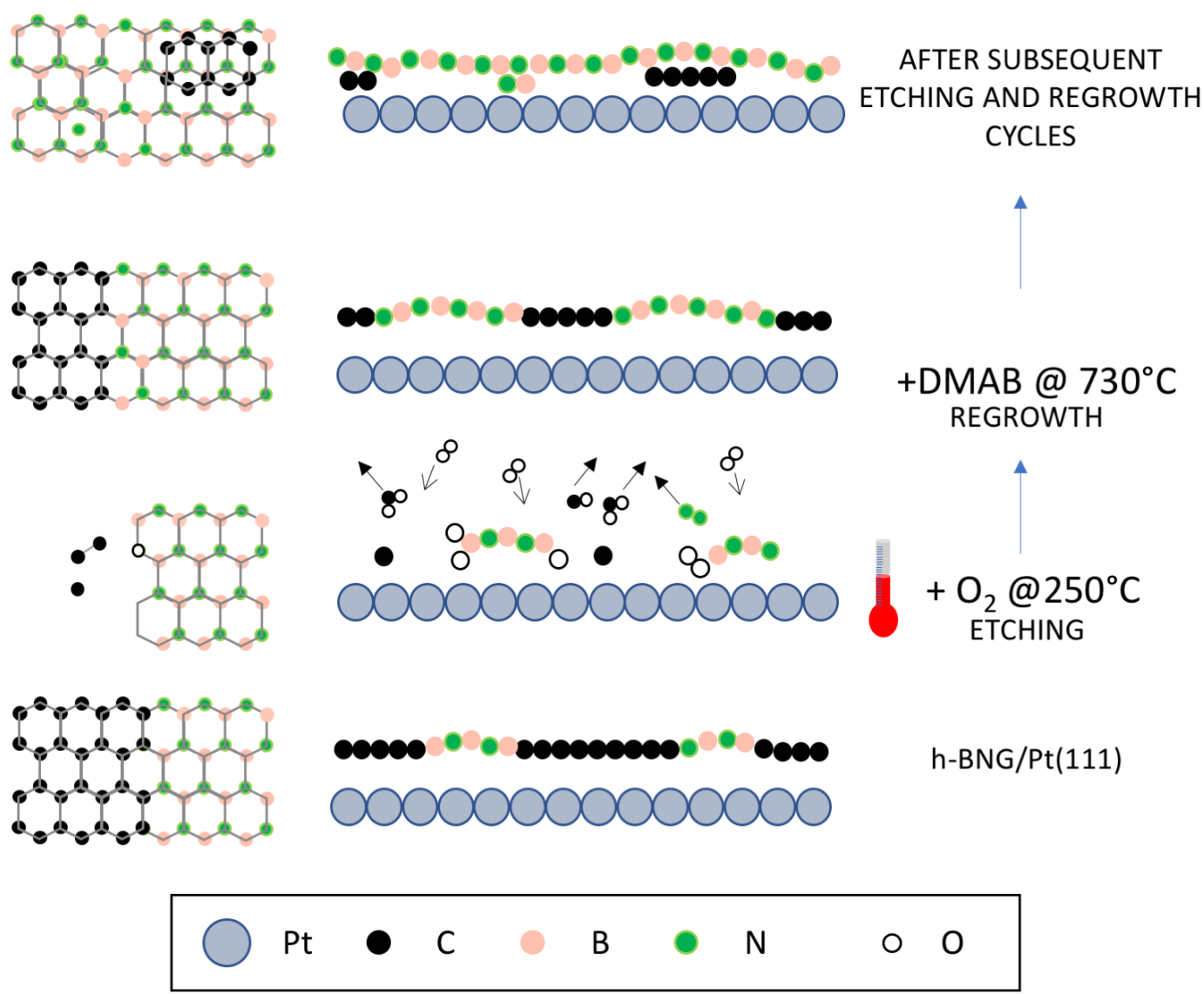

**Figure 6.** Schematic model of the oxygen interaction with h-BNG/Pt(111) at a temperature of 250°C and after subsequent exposures to DMAB precursor at 730° C and oxygen exposures at 250° C. In each panel, left: top view, right: side view.

## 4. Conclusions

We examined the stability and the changes in a heterostructure of hexagonal boron nitride and graphene on platinum in an oxygen environment. Its reactivity against oxygen was evaluated in a wide range of $O_2$ pressures and at temperatures up to 250°C. We have observed a competition between different processes: B–O bond formation, selective oxygen intercalation and selective etching of graphene accompanied by the oxidation of boron.

The study of h-BNG heterostructure, where the behavior of Gr and h-BN can be investigated simultaneously, confirms the much higher stability of h-BN against oxygen etching in comparison to graphene. This supports the superior suitability of h-BN for coating applications. B–O bonds formed during graphene etching can be fully removed by vacuum annealing at T≥730°C in the presence of DMAB, leading to the formation of a different h-BNG layer with higher h-BN content with respect to the starting sample.

This procedure, i.e. subsequent cycles of $O_2$ etching and growth of h-BNG on etched Gr patches, can be exploited to prepare h-BNG layers with controllable and increasing h-BN/Gr composition. After repeated etching/regrowth cycles, the resulting layer predominantly features h-BN with some isolated Gr islands of ~50 nm average size underneath and fewer islands of an h-BN bilayer formed by a mechanism of "confined" growth.

Author Contributions

The manuscript was written through contributions of all authors. All authors have given approval to the final version of the manuscript.

**Data Access**

The data supporting this study are available from the corresponding author upon reasonable request.

Funding Sources

We acknowledge support from MUR (Eurofel project, FOE progetti internazionali).